\documentclass[runningheads]{llncs}
\usepackage[T1]{fontenc}
\usepackage{graphicx}

\usepackage{caption}
\usepackage{subcaption}
\usepackage{multirow}
\usepackage{tabularx}
\usepackage{makecell}
\usepackage{amsmath}
\usepackage{amssymb}

\begin{document}
\title{Sustainability Evaluation Metrics for Recommender Systems}
%
%
\author{Alexander Felfernig\inst{1}\orcidID{0000-0003-0108-3146} \and
Damian Garber\inst{1}\orcidID{0009-0005-0993-0911} \and
Viet-Man Le\inst{1}\orcidID{0000-0001-5778-975X} \and
Sebastian Lubos \inst{1}\orcidID{0000-0002-5024-3786} \and 
Thi Ngoc Trang Tran \inst{1}\orcidID{0000-0002-3550-8352
}
}
\authorrunning{A. Felfernig et al.}
%
\institute{Graz University of Technology, Inffeldgasse 16b/2, 8010 Graz, Austria 
\email{\{alexander.felfernig, damian.garber, v.m.le, slubos, trang.tran\}@tugraz.at}\\
\url{https://www.tugraz.at}}
\maketitle              
\begin{abstract}
Sustainability-oriented evaluation metrics can help to assess the quality of recommender systems beyond wide-spread metrics such as accuracy, precision, recall, and satisfaction. Following the United Nations`s sustainable development goals (SDGs), such metrics can help to analyse the impact of recommender systems   on  environmental, social, and economic aspects. We discuss different basic sustainability evaluation metrics for recommender systems and analyze their applications. 

\end{abstract}
\section{Introduction}\label{sec:introduction}
Recommender systems have become a central component of digital platforms ranging from e-commerce and multimedia to healthcare and education \cite{Alhijawietal2022,JannachetalRecSysOverview2010,RiccietalRecommenderSystems2022}. The success of these systems has been measured using metrics that focus on direct system performance and user engagement such as accuracy, precision, recall, and satisfaction \cite{Jadonetal2025,ZangerleBauer2022}. As the societal impact of these systems grows, there is a need of evaluation criteria that also take into account sustainability aspects including environmental, social, and economic dimensions \cite{Banerjeeetal2024,baniketal2023,Felfernigetal2023SustainabilityRecSys,GeninattiGreenFashion2024,Mauroetal2024,Merinov2023}. Due to the need of taking into account these aspects in recommender systems, \emph{sustainability-aware evaluation metrics} have to be provided \cite{Felfernigetal2023SustainabilityRecSys}. Such  metrics extend the scope of recommender system assessment by embedding long-term  considerations into evaluation processes. These metrics aim not only to optimize the utility of recommendations for individual users or platform operators but also to assure the alignment with sustainability objectives such as the \emph{United Nation`s SDGs}.\footnote{https://sdgs.un.org/goals} By taking into account, for example, the environmental costs of recommended items, the inclusivity of recommendations, and economic equity, these metrics provide a more holistic framework for understanding and improving recommender systems.

In this paper, we provide an overview of basic sustainability-oriented evaluation metrics (numerous further variants thereof can be envisioned) and relate those metrics to the \emph{three sustainability pillars} of the United Nations`s SDGs which are \emph{environmental}, \emph{social}, and \emph{economic}. In this context, we also discuss related applications and indicate directions of future research. With this paper, we want to contribute to a paradigm shift in recommender systems evaluation with a focus on criteria supporting better outcomes for people and the planet.

The remainder of this paper is organized as follows. In Section \ref{sec:environment}, we discuss  evaluation metrics related to the aspect of \emph{environmental sustainability}. Thereafter, in Section \ref{sec:social}, we focus on \emph{social sustainability} evaluation metrics. Section \ref{sec:economic} focuses on \emph{economic sustainability} aspects. Sections \ref{sec:further} and \ref{sec:directions} focus on cross-cutting  metrics and a discussion of open issues for future research. Finally, the paper is concluded with Section \ref{sec:conclusions}.

\section{Environmental Sustainability Metrics}\label{sec:environment}

Environmental sustainability metrics extend the evaluation of recommender systems beyond conventional accuracy-based measures to assess their contribution to environmental goals. 

\subsection{Carbon Footprint of Recommended Items}

The carbon footprint of recommended items measures the average greenhouse gas  emissions (expressed, for example, in tons of $CO_2$ produced within the scope of the whole item lifecycle) associated with the items recommended to users \cite{kalisvaart2025carbonfootprintawarerecommendersystems,SpilloetalCarbonFootprintRecSysUMAP2025,Spilloetal2023}. Let $\mathcal{R}_u$ denote the set of items $\mathcal{R}$ recommended to user $u$,\footnote{This semantics of $\mathcal{R}_u$ also holds for other metrics discussed in this paper.} and $\text{CarF}(i)$ denote the estimated carbon footprint of item $i$. Then, the \emph{average carbon footprint of recommended items} ($\text{AvgCarFI}$)  across all recommender users can be defined as follows:

\vspace{-0.1cm}
\begin{equation}
\text{AvgCarFI} = \frac{1}{|\mathcal{U}|} \sum_{u \in \mathcal{U}} \frac{1}{|\mathcal{R}_u|} \sum_{i \in \mathcal{R}_u} \text{CarF}(i)
\end{equation}

In this context, lower values of $\text{AvgCarFI}$ indicate that the recommender system promotes less carbon-intensive items.

\subsection{Green Item Recommendation Rate}

The green item recommendation rate measures the proportion of recommended items that are labeled as environmentally sustainable \cite{STOCKHEIM2024140157}, for example, items produced with recycled materials, items produced locally, and items certified on the basis of eco labels (e.g., Energy Start\footnote{https://www.energystar.gov/}, EU Ecolabel\footnote{https://environment.ec.europa.eu/}, and Fairtrade\footnote{https://www.fairtrade.net}). Let $\mathcal{G}$ be the set of all green items $i$ in the catalog, and $\mathbb{I}(i)$ be an indicator function that returns 1 if item $i$ is green, and 0 otherwise. Then, the \emph{Green Item Recommendation Rate} (GIRec) can be calculated as follows:

\vspace{-0.1cm}
\begin{equation}
\text{GIRec} = \frac{\sum_{u \in \mathcal{U}} \sum_{i \in \mathcal{R}_u} \mathbb{I}(i)}{\sum_{u \in \mathcal{U}} |\mathcal{R}_u|}
\end{equation}

A higher GIRec implies that the recommender system promotes more environmentally preferable options.

\subsection{Energy Consumption of Recommending}

Energy consumption of \emph{recommending}  refers to the energy consumption required to generate recommendations \cite{Arabzadehetal2024,BeeletalACallForAttention2025,SpilloetalDataReduction2025,Venteetal2024}. Let $E_{\text{inference}}$ denote the total energy (e.g., in kilowatt-hours) consumed by the recommendation system over a defined evaluation period, and $N_{\text{rec}}$ be the total number of recommendations made. The \emph{Energy Consumption per Recommendation} (ECRec) can be defined as follows:

\vspace{-0.1cm}
\begin{equation}
\text{ECRec} = \frac{E_{\text{inference}}}{N_{\text{rec}}}
\end{equation}

This metric is particularly relevant for large-scale deployments utilizing deep learning techniques, where optimization for energy efficiency is crucial.

\subsection{Energy Consumption of Model Building}

Energy consumption of \emph{recommendation model building} refers to the total energy consumed during recommender model building. Let $EC_{build}$ represent the cumulative energy consumed during the complete model construction process, and $N_{epoch}$ denote the number of model learning training epochs performed. The \emph{Energy Consumption per Training Epoch} (ECTrain) can be defined as follows:

\vspace{-0.1cm}
\begin{equation}
\text{ECTrain} = \frac{EC_{build}}{N_{epoch}}
\end{equation}

Model building  can also be evaluated by the volume of training data processed ($N_{dataprocessed}$), one may define the amount of \emph{Energy Consumption per Processed Data Unit} (ECPDat):

\vspace{-0.1cm}
\begin{equation}
\text{ECPDat} = \frac{EC_{build}}{N_{dataprocessed}}
\end{equation}

Such metrics are important for comparing the environmental and computational costs of model architectures or training strategies, especially in scenarios involving frequent recommendation model updates and/or fine-tuning.

\subsection{Energy Savings Through Recommendation}
\emph{Energy Savings through Recommendation} (ESTRec) refers to the reduction in energy consumption or resource usage achieved as a result of applying recommender systems \cite{Borattoetal2024,DeconciniRecSysPersonalValues2024,Felfernigetal2023SustainabilityRecSys,jannach2024recommendersystemsgoodrs4good}. Recommender systems can drive efficiency by optimizing user choices, reducing unnecessary consumption, or streamlining system operations. Let $EC_{baseline}$ denote, for example, the energy consumption of a system (e.g., energy consumption of a household per year) without the use of a recommender, and $EC_{withrec}$ represent the energy consumption observed with the recommender in place (e.g., a recommender for supporting energy savings \cite{WeietalRecommenderEnergySavings2018}). The energy savings can be expressed as:

\vspace{-0.1cm}
\begin{equation}
\text{ESTRec} = \frac{EC_{baseline} - EC_{withrec}}{EC_{baseline}}
\end{equation}

Such metrics are relevant in domains such as smart grids \cite{Felfernigetalinternetofthings2019}, e-commerce \cite{TrangetalSustainabilityExplanationsRecSys2024}, smart homes \cite{Starkeetal2021}, and energy delivery networks \cite{Deconcinietal2025}, where recommendations can influence consumption patterns  \cite{Felfernigetal2023SustainabilityRecSys}.

\subsection{Waste Reduction Through Recommendation}

Recommender systems can contribute to the reduction of  waste by recommending, for example, reusable items \cite{Ichiietal2009} or food \cite{essay101060}. Let $R_{baseline}$ denote the \emph{item reuse rate} of a system without the use of a recommender system, and $R_{withrec}$ represent the reuse rate observed with the recommender in place,  then \emph{Reuse Through Recommendation} (RTR) can be defined as follows:

\vspace{-0.1cm}
\begin{equation}
\text{RTR} = R_{withrec} - R_{baseline}
\end{equation}


Environmental sustainability metrics represent a paradigm shift from a short-term  optimization of user engagement to long-term (ecological) aspects. The practical implementation of these metrics faces challenges such as the availability of reliable carbon footprint data and consistent definitions of item sustainability. 

\section{Social Sustainability Metrics}\label{sec:social}
Social sustainability in recommender systems emphasizes the fair, inclusive, and safe delivery of personalized content \cite{SaidSocialSustainability2025}. 
 These metrics extend traditional ones to ensure that the system's design and outputs support social equity and community well-being.

\subsection{Fairness and Bias Metrics}

Related metrics can assess whether recommendation outcomes are equitably distributed across different demographic groups \cite{WangFairnessMetrics2022}. A simple example of a related aspect is \emph{demographic parity}, which requires that the share of recommended items is similar across sensitive attributes (e.g., gender). Let $G$ denote a set of demographic groups and $P_g(i)$ represents the recommendation probability of item $i$ in group $g \in G$. In this context, a simple demographic parity is given if the following condition is taken into account:

\vspace{-0.1cm}
\begin{equation}
P_g(i) \approx P_{g'}(i) \quad \forall \, g, g' \in G ~~~(g \neq g')
\end{equation}

\subsection{Diversity and Serendipity}

To mitigate filter bubbles and promote exposure to diverse perspectives, recommendation diversity is extremely important \cite{JesseetalIntraListSimilarity2023,LUNARDI2020106771}. \emph{Intra-list diversity} $ListD_u$ for a user $u$ can be measured, for example, on the basis of the average pairwise similarity ($sim$) assuming similarity values in the interval ($0,1$):

\vspace{-0.1cm}
\begin{equation}
\text{ListD}_u = 1 - \frac{\Sigma_{i \in \mathcal{R}_u} \Sigma_{j \in \mathcal{R}_u}  sim(i,j) }{|\mathcal{R}_u| \times (|\mathcal{R}_u|-1)} ~~~(i\neq j)
\end{equation}

where $\text{sim}(i, j)$ denotes the similarity between items $i$ and $j$ (e.g., based on content or metadata). Higher values of $\text{ListD}_u$ indicate more diverse recommendations, i.e., a lower similarity between individual item pairs in a recommendation list  $\mathcal{R}_u$ (the recommendations for user $u$).

Furthermore, \emph{serendipity} ($SER$) is the degree of surprisingness of recommendations $i \in \mathcal{R}_u$ that are still relevant (but unexpected) to the user potentially leading to completely new items and ideas \cite{Mouzhietal2010}:

\vspace{-0.1cm}
\begin{equation}
\text{SER}_u = \frac{\Sigma_{i \in \mathcal{R}_u} \neg con(i,\mathcal{Q}_u) \times rel(i,u)}{|\mathcal{R}_u|}
\end{equation}

In this context, $\mathcal{Q}_u$ is the set of popular or previously seen items by user $u$, $\text{rel}(i, u)$ indicates the relevance of  item $i$ for user $u$, and $\neg con(i,\mathcal{Q}_u)$ returns $1$ if $i$ does not belong to $\mathcal{Q}_u$ ($0$ otherwise).

\subsection{Accessibility and Inclusivity}

Accessibility helps to ensure that recommender systems (and also  recommended items) can be effectively used by users with diverse abilities and backgrounds (represented by different groups $g \in G$). Let $\mathcal{C}$ represent a set of \emph{accessibility criteria} (e.g., understandability of explanations and overall usability for people with visual impairments), and $sat$ a function that indicates to which degree items and/or recommender  user interfaces (represented by the set $Q$) meet these criteria (on an evaluation scale of 0..1). Then, \emph{accessibility} ($ACC_g$) for a group $g \in G$ can be defined as follows:

\vspace{-0.1cm}
\begin{equation}
\text{$ACC_g$} = \frac{\Sigma_{q \in \mathcal{Q}} sat(q,\mathcal{C}, g)}{|\mathcal{Q}|}
\end{equation}

A higher score indicates better accessibility. \emph{Inclusivity} is given if accessibility is equitably fulfilled across different groups $g_i \in G$:  

\vspace{-0.1cm}
\begin{equation}
ACC(g_i) \approx ACC(g_j)  \quad \forall \, g_i, g_j  \in G \quad  (i\neq j)
\end{equation}

\subsection{Harmful Item Exposure}

It is important to understand to which extent the system exposes users to harmful, misleading, or distressing content. Let $\mathcal{H}$ be a set of flagged harmful items (e.g., misinformation, hate speech) and  $harm(i,\mathcal{H})$ be a function that indicates such items. The \emph{Harmful Item Exposure Rate} (HIER) can be defined as:

\vspace{-0.1cm}
\begin{equation}
\text{HIER} = \frac{\sum_{u \in \mathcal{U}} \sum_{i \in \mathcal{R}_u} harm(i,\mathcal{H})}{\Sigma_{u \in \mathcal{U}}|\mathcal{R}_u|}
\end{equation}

Minimizing HIER can support the well-being of users \cite{Hauptmannetal2022wellbeing} and aligns with the goal of ethical content delivery, particularly in health, education, and news recommendation (e.g., in social networks).

\subsection{Health Improvement through Recommendation}

\emph{Health Improvement through Recommendation} (HIRec) refers to the enhancement of individual or population health outcomes which is achieved on the basis of a recommender system. Recommender systems are able to promote healthier behaviors by personalizing suggestions related to diet, exercise, sleep, mental wellness, or medical adherence \cite{FelfernigetalSports2024,Lawoetal2021,Tranetal2018HealthyFood}. 

Let  $\mathcal{M}_{with}$ denote a health outcome metric (e.g., average activity level or body mass index) for users of a recommender system, and $\mathcal{M}_{without}$ represent the same metric for users who did not apply the mentioned recommender system. A corresponding health improvement (over all users) can then be expressed, for example, as follows:

\vspace{-0.1cm}
\begin{equation}
\text{$HIRec$} = \frac{\mathcal{M}_{with} - \mathcal{M}_{without}}{\mathcal{M}_{without}}
\end{equation}

This metric can be helpful in application contexts such as digital health platforms, wellness apps, and preventive healthcare systems.

\section{Economic Sustainability Metrics}\label{sec:economic}

Economic sustainability metrics assess the role of recommender systems in supporting inclusive, resilient, and locally grounded economic ecosystems. 

\subsection{Support for Local Businesses}

This metric quantifies the proportion of recommended items that originate from small or local businesses. Let $\mathcal{L}_u$ denote the set of items from local providers (for user $u$) in the item catalog. The \emph{Local Business Promotion Rate} (LBPR) can be defined as follows:

\vspace{-0.1cm}
\begin{equation}
\text{LBPR} = \frac{\sum_{u \in \mathcal{U}}  |\{i \in \mathcal{R}_u: i \in \mathcal{L}_u\}|}{\sum_{u \in \mathcal{U}} |\mathcal{R}_u|}
\end{equation}

Higher LBPR values suggest that a recommender system  supports a kind of community-level economic development.

\subsection{Long-term Customer Satisfaction}

Rather than optimizing for short-term clicks, economic sustainability favors systems that build lasting customer relationships. Let $sat_u^t$ denote the satisfaction score (e.g., rating, review sentiment, or re-engagement behavior) of user $u$ in time period $t$ (e.g. month $t$). A temporal loyalty metric over a time horizon $T$ (number of time periods, e.g., 12 months) can be defined as:

\vspace{-0.1cm}
\begin{equation}
\text{Loyalty}_u = \frac{1}{T} \sum_{t=1}^{T} sat_u^t
\end{equation}

The average loyalty across all users then can be defined as:

\vspace{-0.1cm}
\begin{equation}
\text{AvgLoyalty} = \frac{1}{|\mathcal{U}|} \sum_{u \in \mathcal{U}} \text{Loyalty}_u
\end{equation}

Sustainable systems aim to maximize $\text{AvgLoyalty}$ rather than focusing solely on short-term engagement metrics. In this context, it could also make sense to give more recent loyalty values a higher weight.

\subsection{Fairness in Exposure}

Recommender systems can inadvertently concentrate exposure and related revenue among a small subset of producers (in a specific item category). To promote economic fairness \cite{JinetalFairnessRecSys2023}, we define fairness in the context of producer exposure\footnote{A similar metric could be \emph{item exposure fairness} (in a specific item category).} (\emph{Producer Exposure Fairness} -- PEF) as follows:

\vspace{-0.1cm}
\begin{equation}
\text{PEF} =  \frac{avgdist_2(\mathcal{P})}{maxdist_2(\mathcal{P})}
\end{equation}

In this context, $avgdist_2(\mathcal{P})$ denotes the average distance between \emph{two} producers in $\mathcal{P}$ and $maxdist_2(\mathcal{P})$ the maximum distance between \emph{two} producers in $\mathcal{P}$ (in terms of the corresponding \emph{exposure count}). 

Economic sustainability metrics provide critical insights into how recommenders affect the distribution of economic value. These metrics help to identify whether a system supports small businesses, encourages longterm user relationships, and promotes equitable economic outcomes. 


\section{Cross-cutting Sustainability Metrics}\label{sec:further}

Cross-cutting sustainability metrics address recommender systems' impacts across environmental, social, and economic dimensions. 

\subsection{Sustainable User Behavior}

This basic metric helps to evaluate sustainability-related user behavior when interacting with a recommender system. Let $B_u$ represent the set (more precisely, the bag) of user behaviors when interacting with the recommender system (e.g., clicking on item details, reading explanations, or purchasing an item). Furthermore, let $\mathcal{S}$ be  a set of sustainable behaviors (e.g., purchasing eco-friendly items). The \emph{Sustainable Behavior Score} (SBS) can be defined as:

\vspace{-0.1cm}
\begin{equation}
\text{SBS} = \frac{\Sigma_{ u \in U} |\{b \in B_u: b \in \mathcal{S}\}|}{\Sigma_{ u \in U} |B_u|}
\end{equation}

Higher SBS values indicate a higher degree of sustainability-related user behaviors when interacting with a recommender system.

\subsection{Interpretability of Recommendations}

Interpretability ($IntP$) of recommendations is crucial for enabling informed decision-making \cite{Deconcinietal2025}. Let $\mathcal{E}_u$ denote the set of explanations $e$ shown to user $u$ and let $\text{interpret}(e)$ measure the interpretability of an explanation. The \emph{Average Explanation Interpretability} (IntP) across all users is:

\vspace{-0.1cm}
\begin{equation}
\text{IntP} = \frac{1}{|\mathcal{U}|} \sum_{u \in \mathcal{U}} \frac{1}{|\mathcal{E}_u|} \sum_{e \in \mathcal{E}_u} \text{interpret}(e)
\end{equation}

Interpretability can be estimated, for example, on the basis of user feedback, readability heuristics, information complexity measures, and potentially also from an LLM-based understandability analysis.

\subsection{Life Cycle Impact of Recommendations}

Life cycle impact analysis includes upstream and downstream effects in the context of production, sales, usage, and disposal of recommended items. Let $\text{LCI}(i)$ denote the total estimated life cycle impact score for item $i$ which could include different aspects such as carbon footprint but beyond that also aspects such as impacts of harmful content delivery and impact of item reuse. The \emph{Average Life Cycle Impact of Recommendations} (AvgLCI) is:

\vspace{-0.1cm}
\begin{equation}
\text{AvgLCI} = \frac{1}{|\mathcal{U}|} \sum_{u \in \mathcal{U}} \frac{1}{|\mathcal{R}_u|} \sum_{i \in \mathcal{R}_u} \text{LCI}(i)
\end{equation}

A lower $\text{AvgLCI}$ indicates that the system promotes items with lower environmental and social burdens over their life cycles.

The practical deployment of these cross-cutting  metrics also depends on the generation and accessibility of  trustworthy metadata. 

\section{Challenges and Research Directions}\label{sec:directions}

Despite a growing interest in sustainability-aware recommender systems \cite{Felfernigetal2023SustainabilityRecSys,jannach2024recommendersystemsgoodrs4good}, several critical challenges hinder the widespread adoption and standardized evaluation with sustainability metrics. 

\subsection{Multi-objective Optimization}

Incorporating sustainability goals often introduces trade-offs between traditional performance metrics (e.g., precision or click-through rate) and sustainability outcomes. Formally, we want to optimize a vector-valued objective function:

\vspace{-0.1cm}
\begin{equation}
\max_{\theta} \quad \mathbf{F}(\theta) = [\text{Accuracy}(\theta), \text{Sustainability}(\theta)]
\end{equation}

where $\theta$ are the model parameters. This requires to solve related multi-objective optimization problems.

\subsection{Data Availability and Labeling}

Most sustainability metrics require granular metadata such as carbon footprints, ethical sourcing indicators, or vendor type which is often missing or inconsistently labeled. Let $\mathcal{I}$ be the item set and $s_i$ define a binary sustainability label for items $i \in \mathcal{I}$. Then, the share of labeled items is:

\vspace{-0.1cm}
\begin{equation}
\text{LabelCoverage} = \frac{|\{i \in \mathcal{I} : s_i \text{ is known}\}|}{|\mathcal{I}|}
\end{equation}

A low $LabelCoverage$ reduces the feasibility and accuracy of sustainability-aware evaluation.

\subsection{Benchmarking}

Currently, there is no agreed-upon benchmark dataset or unified metric suite for evaluating recommender systems on sustainability dimensions. This lack of standardization makes cross-system comparisons difficult and limits reproducibility. Future work should prioritize the development of shared datasets, evaluation protocols, and toolkits for sustainable recommendation research.

\section{Productive Usage of Sustainability Metrics}

Different aspects regarding the application of the presented metrics must be taken into account. (1) developers should incorporate sustainability indicators (e.g., carbon footprint) into model evaluation pipelines. This can be done, for example,  by extending  logging frameworks to collect metadata (e.g., energy use, item origin, and user demographics) and enabling sustainability-aware performance dashboards. (2) recommender algorithms can be optimized based on sustainability aspects. For example, on the basis of  "green item filters" for the purpose of prioritizing items. Furthermore, companies can  report  sustainability performance of their recommendation engines.



\section{Conclusions}\label{sec:conclusions}

Sustainability-oriented evaluation metrics are critical for evolving recommender systems beyond conventional metrics. By incorporating environmental, social, and economic considerations, these metrics align recommender system development and evaluation with global sustainability objectives, particularly the United Nations Sustainable Development Goals (SDGs). Through the promotion of eco-friendly products, equitable access, and support for local economies, recommender systems can play a transformative role in fostering responsible digital consumption. With this paper, we contribute to the goal of explicitly integrating sustainability aspects into recommender system evaluation -- particularly, by proposing a basic core set of related evaluation metrics. 



\bibliographystyle{plain}
\bibliography{bibliography}

\end{document}